\documentclass{Interspeech2024}
\usepackage{booktabs}
\usepackage{multirow}
\usepackage[table,xcdraw]{xcolor}
\usepackage{booktabs}
\usepackage{subcaption}
\interspeechcameraready

\title{MM-KWS: Multi-modal Prompts for Multilingual User-defined \\ Keyword Spotting}

\name[]{Zhiqi}{Ai}
\name[]{Zhiyong}{Chen}
\name[]{Shugong}{Xu}

\address{
  School of Communication and Information Engineering, Shanghai University, Shanghai, China
}
\email{\{aizhiqi-work, zhiyongchen, shugong\}@shu.edu.cn}

\keywords{user-defined keyword spotting, multi-modal, multilingual, hard case mining, zero-shot learning}

\begin{document}

\maketitle
\begin{abstract}

In this paper, we propose MM-KWS, a novel approach to user-defined keyword spotting leveraging multi-modal enrollments of text and speech templates. Unlike previous methods that focus solely on either text or speech features, MM-KWS extracts phoneme, text, and speech embeddings from both modalities. These embeddings are then compared with the query speech embedding to detect the target keywords. To ensure the applicability of MM-KWS across diverse languages, we utilize a feature extractor incorporating several multilingual pre-trained models. Subsequently, we validate its effectiveness on Mandarin and English tasks. In addition, we have integrated advanced data augmentation tools for hard case mining to enhance MM-KWS in distinguishing confusable words. Experimental results on the LibriPhrase and WenetPhrase datasets demonstrate that MM-KWS outperforms prior methods significantly.

\end{abstract}

\section{Introduction}

Traditional keyword spotting (KWS) systems typically rely on extensive datasets for recognizing predefined keywords like "Ok Google" and "Hey Siri" \cite{higuchi20_interspeech, lopez2021deep}. However, customizing these keywords requires costly data collection and model training processes. To address this issue, recent research has shifted focus towards user-defined keyword spotting (UDKWS) \cite{wang21ea_interspeech, huang2021query, baseline_cmcd, baseline_emkws}, which aims to accurately detect new keywords using limited examples, thus offering a more flexible and efficient solution to meet users' needs.

Previous studies have utilized large vocabulary continuous speech recognition (LVCSR) systems to transcribe audio into lattices, facilitating keyword search and achieving high accuracy \cite{motlicek2012improving, chen2013quantifying}. However, these systems are limited by their predefined vocabularies, which results in decreased performance when encountering out-of-vocabulary keywords \cite{huang2021query}. Recent advancements have attempted to address this limitation by incorporating hotword lists into automatic speech recognition (ASR) \cite{yao21_interspeech, funasr1}, thereby enhancing recall and accuracy for target keywords. Despite these improvements, such approaches often incur high costs, making them less viable for deployment in low-resource scenarios.

Currently, most UDKWS systems employ the query-by-example (QbyE) approach \cite{wang21ea_interspeech, huang2021query, kim2019query}. Some studies allow users to register speech templates for customized keywords, called QbyA. \cite{wang21ea_interspeech, kirandevraj2022generalized} conducted frame-level embedding extraction using a pre-trained acoustic model and utilized dynamic time warping (DTW) for measuring similarity between the registration embedding and the query embedding. \cite{huang2021query, lee2023fully} focused on studying acoustic word embedding to significantly reduce matching cost. However, the QbyA method's performance is heavily dependent on recording consistency and entails a laborious registration process \cite{baseline_cmcd}. Therefore, recent research emphasizes cross-modal matching of text and speech modalities for customized keywords, known as QbyT. \cite{baseline_cmcd} leverages text-speech correspondence to achieve excellent performance. \cite{baseline_emkws} and \cite{baseline_phonmatchnet} have optimized the matching scheme and loss function, respectively, resulting in higher performance gains. \cite{baseline_adakws} achieves the current lowest Equal Error Rate (EER) through larger-scale pre-training and negative mining. Despite QbyT's reliability and user-friendly registration process, challenges arise, especially for speakers with accents. Mispronunciations of phonemes can significantly affect performance \cite{baseline_ced}.

To tackle these challenges, we present MM-KWS, a novel approach to UDKWS leveraging multi-modal enrollments of text and speech templates. To ensure MM-KWS's adaptability across various languages, we employ a feature extractor comprising several multilingual pre-trained models. To validate MM-KWS's effectiveness in Mandarin, we introduce the WenetPhrase dataset, filling the gap in existing Mandarin KWS datasets by providing evaluation data for confusable words. Furthermore, we integrate advanced data augmentation tools for hard case mining to bolster MM-KWS's capability to distinguish confusable words. As a result, MM-KWS outperforms previous approaches on both LibriPhrase and WenetPhrase datasets and exhibits outstanding zero-shot performance. The implementation code of our proposed model and WenetPhrase dataset are available at Project page\footnote{https://github.com/aizhiqi-work/MM-KWS}.  Our main contributions are as follows:

\begin{itemize}
    \item We propose MM-KWS, a novel multi-modal prompts for user-defined keyword spotting method that utilizes text and speech templates as multi-modal enrollments.
    \item We introduce several multilingual pre-trained models to the keyword spotting task efficiently and validate the high performance of MM-KWS on English and Mandarin data.
    \item We employ advanced data augmentation tools for hard case mining in keyword spotting tasks, thereby strengthening MM-KWS's capability to distinguish confusable keywords.
\end{itemize}

\begin{figure*}[t]
  \centering
  \includegraphics[width=0.9\linewidth]{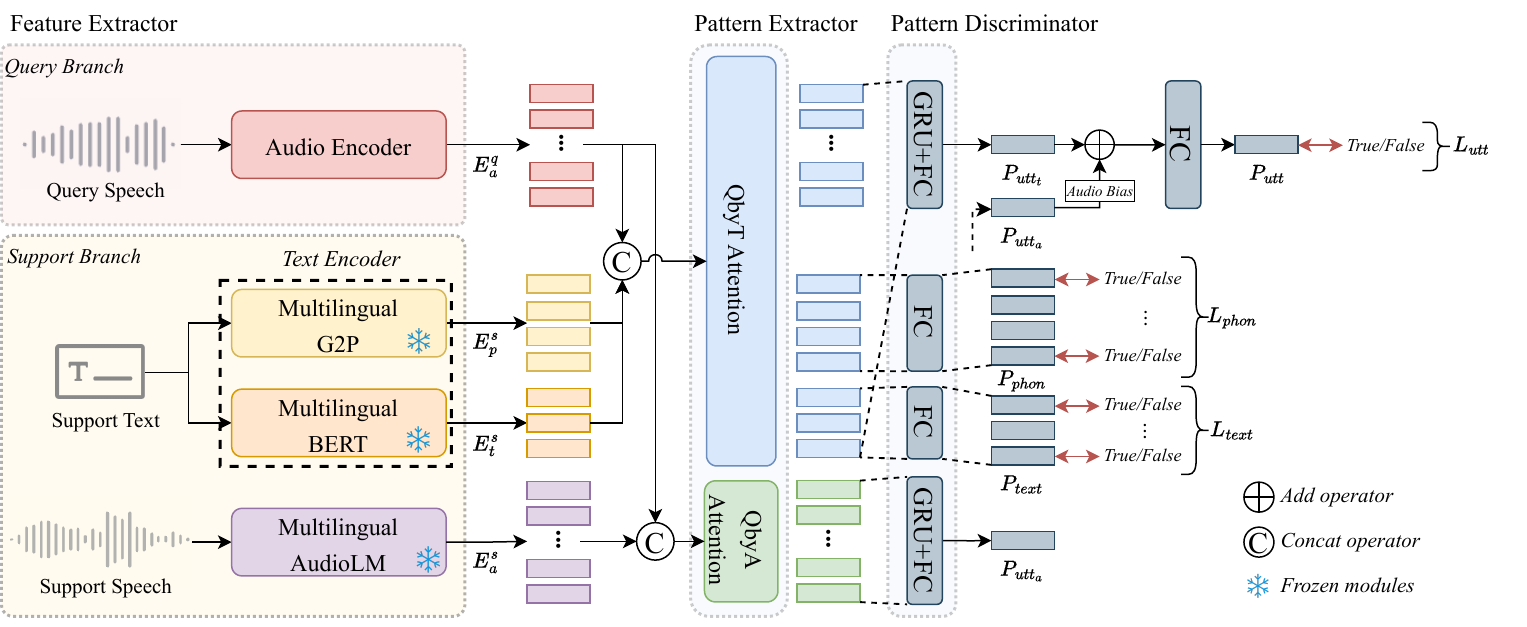}
  \caption{Overall architecture of the proposed model, MM-KWS. Multi-modal information by extracting embeddings from enrolled text and speech templates to match with query speech embeddings.}
  \label{fig:kws-train-pipeline}
\end{figure*}

\section{Proposed Method}
In this section, we introduce our proposed model, Multi-modal Prompts for User-defined Keyword Spotting (MM-KWS). MM-KWS consists of three sub-modules: a feature extractor, a pattern extractor, and a pattern discriminator. The overall architecture is shown in Figure \ref{fig:kws-train-pipeline}.

\subsection{Architecture}
\textbf{Feature Extractor:}
The feature extractor module comprises a query branch and a support branch, as illustrated in Figure \ref{fig:kws-train-pipeline}. Inspired by \cite{baseline_emkws, baseline_ced}, we adopt the Conformer architecture \cite{conformer} as the audio encoder to transform the query speech into the speech embedding. The support branch comprises a dual-branch text feature extractor and a high-performance speech encoder. These components extract phoneme, text, and speech embeddings from text and speech templates, respectively.

Through the feature extractor module, we represent the query speech embedding as \( E^q_{a} \in \mathbb{R}^{T^{q}_{a} \times d} \), the support phoneme embedding as \( E^s_{p} \in \mathbb{R}^{T^s_{p} \times d} \), the support text embedding as \( E^s_{t} \in \mathbb{R}^{T^s_{t} \times d} \), and the support speech embedding as \( E^s_{a} \in \mathbb{R}^{T^s_{a} \times d} \), where \( T^q_{a} \) represents the query speech frame length. \( T^s_{p} \) and \( T^s_{t} \) represent the number of phonemes and subwords in the support text, respectively. \( T^s_{a} \) represents the length of the support speech frame, and \( d \) represents the frame dimension.

\textbf{Pattern Extractor:}
The pattern extractor is built upon the self-attention mechanism. As elucidated in \cite{baseline_phonmatchnet}, the self-attention method yields commendable cross-modal matching performance, particularly exhibiting efficiency in KWS. The pattern extractor comprises a query-by-text attention module (QTAM) and a query-by-audio attention module (QAAM).

The QTAM module processes three inputs: the query speech embedding \(E^q_{a}\), the support phoneme embedding \(E^s_{p}\), and the support text embedding \(E^s_{t}\). To differentiate features derived from these sources, we use three learnable coding vectors \(e^{type}\), each indicating the source of the features. Temporal position encoding, denoted as \(e^{pos}\) and following sinusoidal position encoding, is also incorporated. This results in the transformer inputs as shown in Equation (\ref{eq:eqt}).
\begin{equation}
    \overline{E}= E + e^{pos} + e^{type}
    \label{eq:eqt} 
\end{equation}

Subsequently, the QTAM is utilized to conduct cross-modal matching between text and speech modalities, specifically determining whether the query speech corresponds to the target text. The transformed features $\overline{E^q_{a}}$, $\overline{E^s_{p}}$, and $\overline{E^s_{t}}$ are concatenated along the temporal dimension as $E^c_{ta}$, and the joint features $E^{j}_{ta}$ are computed using self-attention:
\begin{align}
    E^c_{ta} &= (\overline{E^q_{a}};\overline{E^s_{p}};\overline{E^s_{t}}) \in \mathrm{R}^{(T^q_{a}+T^s_{p}+T^s_{t}) \times d} \\
    E^{j}_{ta} &= \text{Attention}(E^c_{ta}, E^c_{ta}, E^c_{ta}) \in \mathrm{R}^{(T^q_{a}+T^s_{p}+T^s_{t}) \times d}
\end{align}

The QAAM takes two specific inputs: the query speech embedding \( E^q_{a} \) and the support speech embedding \( E^s_{a} \). The QAAM is utilized to determine if the query speech corresponds to the enrolled speech templates. To distinguish between embeddings from these sources, we apply similar learnable type coding vectors and position encoding as outlined in Equation (\ref{eq:eqt}). The joint features \( E^{j}_{aa} \) are calculated using the defined attention mechanism:
\begin{align}
E^c_{aa} &= (\overline{E^q_{a}}; \overline{E^s_{a}}) \in \mathrm{R}^{(T^q_{a}+T^s_{a}) \times d} \\
E^{j}_{aa} &= \text{Attention}(E^c_{aa}, E^c_{aa}, E^c_{aa}) \in \mathrm{R}^{(T^q_{a}+T^s_{a}) \times d}
\end{align}

\textbf{Pattern Discriminator:}
For the pattern discriminator, we employ the GRU module to derive the respective utterance-level posterior probabilities from the joint embedding of QTAM and QAAM, denoted as \( P_{utt_{t}} \) and \( P_{utt_{a}} \), respectively. MM-KWS decision primarily relies on the more stable output from support text, with support speech providing supplementary information. Thus, the decision output is expressed as \(P_{utt_t}\), enhanced by \(P_{utt_a}\) to jointly assess utterance-level matching.

\begin{equation}
P_{utt} = \sigma (W^{u} \cdot (P_{utt_t} + P_{utt_a}) + b^{u})
\end{equation}

In addition, we computed the match probabilities for phonemes $P_{phon}$ and text $P_{text}$ from the phoneme sequences $(E^{j}_{ta})_{phon}$ and word sequences $(E^{j}_{ta})_{text}$ in the joint embedding of QTAM, respectively. The subscripts 'phon' and 'text' indicate frame indices in the ranges ($T^q_{a}$, $T^q_{a}+T^s_{p}$] and ($T^q_{a}+T^s_{p}$, $T^q_{a}+ T^s_{p}+T^s_{t}$], respectively.

\begin{table}[t]
\caption{Experimental results of the proposed MM-KWS model on the Libriphrase dataset compared to the baseline model. (*): our model with confusable keywords generation.}
\resizebox{\linewidth}{!}{
\begin{tabular}{@{}lcccccc@{}}
\toprule
\multicolumn{1}{l}{\multirow{2}{*}{Method}} & \multirow{2}{*}{\# Params} & \multicolumn{2}{c}{AUC(\%) ↑} & \multicolumn{1}{c}{} & \multicolumn{2}{c}{EER(\%) ↓} \\ \cmidrule(lr){3-4} \cmidrule(l){6-7} 
\multicolumn{1}{c}{} &                                & LH       & LE       &  & LH        & LE        \\ \midrule
Whisper-Tiny \cite{baseline_whisper}       & 39M      & 73.37           & 89.19             &  & 33.04          & 17.31     \\
Whisper-Small \cite{baseline_whisper}      & 224M     & 82.90           & 95.92             &  & 21.45          & 8.14      \\
Whisper-Large \cite{baseline_whisper}      & 1550M    & 85.80           & 97.54             &  & 19.57          & 5.33      \\ \midrule
Triplet \cite{baseline_triplet}            & N/A      & 54.88           & 63.53             &  & 44.36          & 32.75     \\
% Attention \cite{baseline_attention}        & N/A      & 62.65           & 78.74             &  & 41.95          & 28.74     \\
% DONUT \cite{baseline_donut}                & N/A      & 54.88           & 63.53             &  & 44.36          & 32.75     \\
CMCD \cite{baseline_cmcd}                  & 0.7M     & 73.58           & 96.70             &  & 32.90          & 8.42      \\
% CLAD \cite{baseline_clad}                  & 2.2M     & 76.15           & 97.03             &  & 30.30          & 8.65      \\ 
EMKWS \cite{baseline_emkws}                & 3.7M     & 84.21           & 97.83             &  & 23.36          & 7.36      \\
PhonMatchNet \cite{baseline_phonmatchnet}  & 0.7M     & 88.52           & 99.29             &  & 18.82          & 2.80      \\
CED \cite{baseline_ced}                    & 3.6M     & 92.70           & 99.84             &  & 14.40          & 1.70      \\
AdaKWS-Tiny \cite{baseline_adakws}         & 15M      & 93.75           & 99.80             &  & 13.47          & 1.61      \\
AdaKWS-Small \cite{baseline_adakws}        & 109M     & 95.09           & 99.82             &  & 11.48          & 1.21      \\ \midrule
MM-KWS                                     & 3.9M     & 94.02  & \textbf{99.98}    &  & 12.45 & \textbf{0.41}      \\ 
MM-KWS*                                    & 3.9M     & \textbf{96.25}  & 99.95    &  & \textbf{9.30}  & 0.68      \\ \bottomrule
\end{tabular}
}
\label{tab:LibriPhrase}
\end{table}

\subsection{Training Approach}
Our training objective is denoted by $\mathcal{L}_{total}$, which consists of a combination of three binary cross-entropy (BCE) losses. These include the utterance-level loss ($\mathcal{L}_{utt}$) as the main loss, and two auxiliary losses: the phoneme-level detection loss ($\mathcal{L}_{phon}$) and the word-level detection loss ($\mathcal{L}_{text}$).
\begin{equation}
    \mathcal{L}_{total} = \mathcal{L}_{utt} + \mathcal{L}_{phon} + \mathcal{L}_{text}
\end{equation}

In evaluating the similarity between the query branch and support branch as a whole ($\mathcal{L}_{utt}$), we assign a label of 1 if the query speech is the target keyword; otherwise, the label is 0. Additionally, $\mathcal{L}_{phon}$ and $\mathcal{L}_{text}$ are utilized to assess whether the target unit is present in the query speech. A label of 1 is assigned if the query speech contains the phonemes or words of the support text; otherwise, the label is 0.

\subsection{Data Augmentation with Advanced Tools}
To enhance the robustness of MM-KWS against confusable words, we leverage advanced data augmentation tools, which involve confusable keyword generation (Stage 1) and speech synthesis (Stage 2).

In Stage 1, our objective is to acquire words susceptible to confusion, primarily involving semantically or phonetically similar words. Initially, we employ a rule-based approach, leveraging a pre-trained grapheme-to-phoneme (G2P) \cite{baseline_phonmatchnet, park20c_interspeech} model and DistilBERT \cite{sanh2019distilbert} to convert a corpus of common words in the target language into phoneme sequences and semantic embeddings, respectively. Subsequently, we compute edit distances on phoneme sequences to identify words phonetically similar to the target word, thus establishing closely related terms. To pinpoint words with similar semantics, we calculate the cosine similarity between their semantic embeddings and those of the target word. Additionally, through word permutation, we generate complementary negative instances. Furthermore, we harness a large language model to produce a substantial corpus of negative instances of the target word that could occur in real-world scenarios.

In Stage 2, given the absence of corresponding audio data for the generated confused words, we employed a multilingual ZS-TTS\footnote{https://great-research.github.io/tsct-tts-demo/} \cite{chen2024optimizing} for speech synthesis. This study involved the generation of a total of 1.5 million English data and 2.4 million Mandarin data for the training of MM-KWS.

\section{Experiments}
\label{sec:exp}
We implemented our method using PyTorch and conducted experiments on x86 Linux machines equipped with 4 NVIDIA 4090 GPUs.
% \subsection{Datasets and Metrics}
\subsection{Datasets}
\begin{itemize}
\item \textbf{LibriPhrase:} We constructed LibriPhrase in accordance with \cite{baseline_cmcd, baseline_phonmatchnet}. The training dataset is derived from train-clean-100/360, while the test dataset is extracted from train-others-500. The test dataset is further divided into two parts: LibriPhrase Easy (LE) and LibriPhrase Hard (LH). Evaluation metrics primarily include Equal Error Rate (EER) and Area Under the ROC Curve (AUC).

\item \textbf{WenetPhrase:} We introduce the WenetPhrase dataset to evaluate MM-KWS for Mandarin at the scale of LibriPhrase. Speech segmentation was conducted using a forced alignment algorithm \cite{pratap2023scaling} on approximately 1000 hours of WenetSpeech M/S data \cite{zhang2022wenetspeech}. Text tokenization was performed using Jieba to obtain target word lists. Subsequently, speech segments with durations ranging from 0.5 to 2 seconds and containing between 2 and 6 words were selected. This process yielded approximately 122 K training classes and 54 K test classes, for a total of 2.9 M samples. We constructed WenetPhrase Easy (WE) and WenetPhrase Hard (WH) subsets. Examples of the WenetPhrase dataset is shown in Table \ref{tab:subtable1}. Evaluation metrics include EER and AUC.

\item \textbf{SPC:} We assess the zero-shot performance of MM-KWS employing the Speech Commands dataset (SPC) \cite{warden2018speech}. Follow the settings in \cite{lee2023fully}, we conduct evaluation on multi-classification tasks. Specifically, 10 target keywords are randomly chosen, while the remaining 20 keywords are designated as unknown classes. We utilize Acc(close) and Acc(open) to measure multi-classification performance, where Acc(close) does not contain unknown classes.
\end{itemize}

\begin{table}[t]
    \centering
    \caption{Data samples and benchmark on WenetPhrase.}
    \begin{subtable}[t]{\linewidth}
        \centering
        \caption{Examples of WenetPhrase.}
        \includegraphics[width=\linewidth]{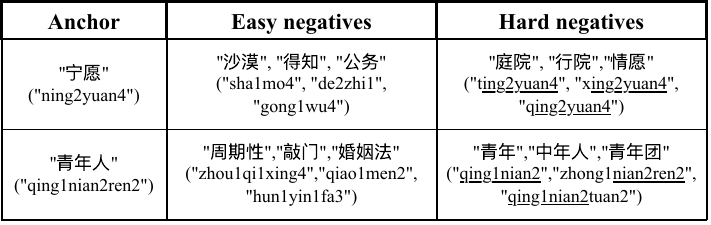}
        \label{tab:subtable1}
    \end{subtable}
    \hspace{0.05\textwidth}
    \begin{subtable}[t]{\linewidth}
        \centering
        \caption{Experimental results of the proposed MM-KWS model on WenetPhrase dataset compared to the baseline model.}
        \resizebox{\linewidth}{!}{
        \begin{tabular}{@{}lccccccc@{}}
        \toprule
        \multicolumn{1}{l}{\multirow{2}{*}{Method}} &
          \multirow{2}{*}{\# Params} &
          \multicolumn{2}{c}{AUC(\%) ↑} &
           &
          \multicolumn{2}{c}{EER(\%) ↓} &
          \multirow{2}{*}{\begin{tabular}[c]{@{}c@{}}Latency \\ (MS)↓\end{tabular}} \\ \cmidrule(lr){3-4} \cmidrule(lr){6-7}
        \multicolumn{1}{c}{}                     &          & WH     & WE               &  & WH     & WE                &                       \\ \midrule
        Whisper-Tiny \cite{baseline_whisper}     & 39M      & 56.53  & 60.67            &  & 44.66  & 45.38             & 102                    \\
        Whisper-Small \cite{baseline_whisper}    & 244M     & 57.31  & 72.20            &  & 44.53  & 35.56             & 183                   \\
        Whisper-Large \cite{baseline_whisper}    & 1550M    & 56.46  & 88.77            &  & 48.76  & 15.51             & 316                   \\
        FunASR$\dagger$ \cite{funasr1}          & 220M     & 58.31  & 99.02            &  & 45.03  & 3.62              & 300                   \\ \midrule
        MM-KWS                              & 3.9M     & 83.73 & \textbf{99.79}   &  & 23.88     & \textbf{1.95}     & 6     \\ 
        MM-KWS*                             & 3.9M     & \textbf{85.84} & 99.15 &  & \textbf{22.06} & 4.25                     & 6                     \\ \bottomrule
        % MM-KWS-Base*$\dagger$                    & 3.9M     & $\sim$ & $\sim$ &  & $\sim$ & $\sim$ & 6      \\ \bottomrule
        \end{tabular}
            }
        \label{tab:WenetPhrase}
    \end{subtable}
\end{table}

\subsection{Training Details}
\label{exp:pmodal}

We utilize Tiny Conformer \cite{conformer} as the audio encoder in the query branch, as depicted in Figure \ref{fig:kws-train-pipeline}. The Conformer architecture is configured with \textit{\{6 encoder layers, encoder dimension of 128, convolution kernel of size 3, and 4 attention heads\}}. In the support branch, we employ multilingual DistilBERT \cite{sanh2019distilbert} to derive 768-dim text embedding, use multilingual G2P \cite{baseline_phonmatchnet, park20c_interspeech} for converting text into 64-dim phoneme embedding, and leverage the 18-layer high-performance multilingual XLR-S(0.3B) \cite{conneau21_interspeech, 9801640} to extract a 1024-dim speech embedding from the speech templates. Notably, all supported modules are fixed-parameter and are pre-fetched only once during inference, thereby eliminating additional computational costs. We incorporate three lightweight mappers to standardize all inputs into a unified 128-dim embedding. Both Attention modules in the pattern extractor utilize \textit{\{2 encoder layers and 4 attention heads\}}. The output dimension of GRU is 64-dim. For the training process, we employ the Adam optimizer, with 40k training steps.

\begin{figure}[t]
  \centering
  \includegraphics[width=0.95\linewidth]{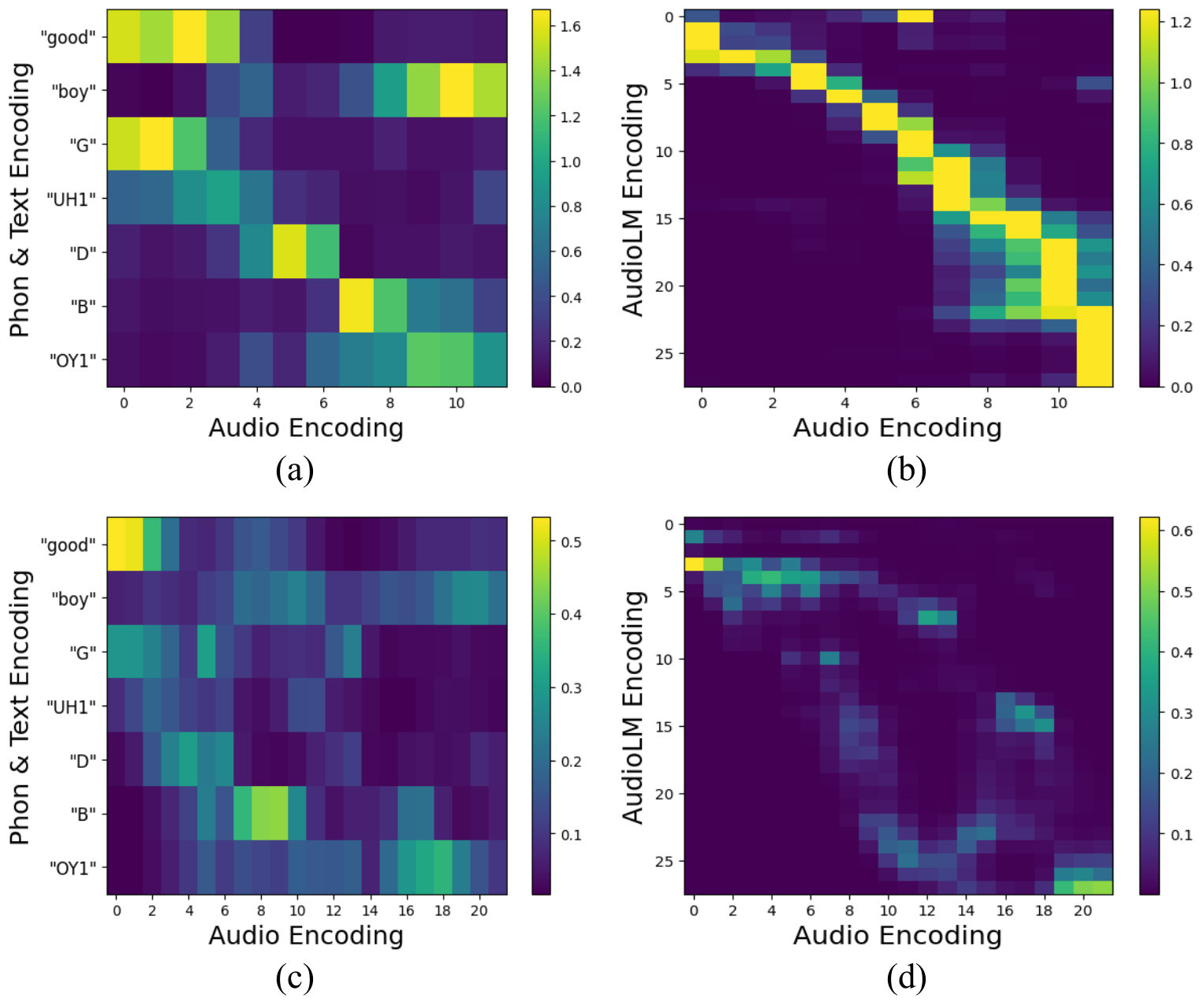}
  \caption{Visualization of pattern extractor attention maps. The target keyword is "good boy". (a) and (b) show positive examples, while (c) and (d) show negative examples, "in the United States."}
  \label{fig:heatmap}
\end{figure}

\section{Results}
\subsection{Comparative Evaluation of MM-KWS}
Table \ref{tab:LibriPhrase} presents a comparison between MM-KWS and previous state-of-the-art methods on the LibriPhrase dataset. The results indicate significant performance improvements with MM-KWS, particularly on LH subset. Notably, MM-KWS*, trained using confusable keywords generation, exhibits an AUC score of 96.25\% and an EER of 9.30\% on LH, surpassing that of AdaKWS \cite{baseline_adakws} with larger parameters and pre-trained on a more extensive dataset.

Similarly, we assess the performance of MM-KWS and state-of-the-art ASR models on the WenetPhrase dataset, as shown in Table \ref{tab:WenetPhrase}. We observe that the difficulty level of WE is comparable to that of LE. Whisper-large \cite{baseline_whisper} demonstrates commendable performance, and FunASR$\dagger$ \cite{funasr1} model equipped with a hot word list achieves better results on WE. However, these ASR systems encounter challenges in processing WH data, suggesting that identifying confusing words in Mandarin poses greater difficulty. Notably, MM-KWS exhibits strong performance on WenetPhrase, achieving an AUC of 99.79\% and an EER of 1.95\% on WE. With data augmentation, MM-KWS* attains an AUC of 85.84\% and an EER of 22.06\% on WH. compared to these ASR systems, MM-KWS demonstrates only a 6ms latency.

\subsection{The Zero-Shot Performance of MM-KWS}

We employ SPC to assess the zero-shot performance of MM-KWS in multi-classification tasks. The baseline system adopts a query-by-audio approach \cite{lee2023fully}. Noteworthy is the subpar performance demonstrated by the QbyA-baseline when dealing with a restricted number of registered speech inputs. In contrast, MM-KWS demonstrates outstanding performance solely with text input, and its effectiveness is further heightened by multi-modal enrollments.

\begin{table}[]
\centering
\caption{The zero-shot performance of MM-KWS on SPC.}
\resizebox{0.95\linewidth}{!}{
\begin{tabular}{@{}lccc@{}}
\toprule
\multirow{2}{*}{Method} & \multirow{2}{*}{\# of supports} & \multicolumn{2}{c}{SPC} \\ \cmidrule(l){3-4} 
 &  & Acc(close)↑ & Acc(open)↑ \\ \midrule
\multirow{2}{*}{QbyA-baseline \cite{lee2023fully}} 
& 1 & 69.0±1.67 & 66.0±1.03 \\
& 5 & 90.5±0.53 & 80.6±0.44 \\ \midrule
\multirow{2}{*}{MM-KWS} 
& 0 & \underline{94.4}±0.18 & \underline{88.4}±0.28 \\
& 1 & \textbf{95.2}±0.17 & \textbf{90.6}±0.19\\ \bottomrule
\end{tabular}
}
\label{tab:gmany}
\end{table}

\subsection{Visualization of pattern extractor attention maps}
We visualized the attention maps of the two attention modules in the pattern extractor to investigate the fundamental principles of MM-KWS for discriminating keywords. Figure \ref{fig:heatmap} illustrates these visualizations: (a) depicts the response of phoneme and text embedding along with the query speech embedding during positive sample discrimination, while (b) illustrates the response of registered speech embedding and query speech embedding. In contrast, (c) and (d) represent the scenarios during discrimination of negative samples. It is evident that all three patterns demonstrate clear monotonicity when the test speech contains the target keyword. Conversely, when the test speech is a negative example, such monotonicity is absent. This observation highlights that the pattern extractor aligns query speech at different levels to determine target keywords.

\subsection{Ablation studies of MM-KWS}
Table \ref{tab:ablation} presents our ablation study results. Experiments prove that adding confusable keywords generation, support speech branch, and auxiliary loss all help make MM-KWS better.

\begin{table}[!ht]
\caption{Ablation studies of MM-KWS.}
\centering
\resizebox{\linewidth}{!}{
\begin{tabular}{@{}llccccc@{}}
\toprule
\multirow{2}{*}{Method}            &  & \multicolumn{2}{c}{AUC(\%) ↑} &  & \multicolumn{2}{c}{EER(\%) ↓} \\ \cmidrule(lr){3-4} \cmidrule(l){6-7} 
                                   &  & LH            & LE                  &  & LH            & LE                 \\ \midrule
MM-KWS                             &  & \textbf{96.25}& \underline{99.95}   &  & \textbf{9.30} & \underline{0.68}   \\
w/o confusable keywords generation &  & 94.02         & \textbf{99.98}      &  & 12.45         & \textbf{0.41}      \\
w/o support speech branch          &  & 95.36         & 99.94               &  & 10.41         & 0.82               \\
w/o auxilary loss                  &  & 93.48         & 99.89               &  & 12.95         & 1.35               \\ \bottomrule
\end{tabular}
}
\label{tab:ablation}
\end{table}
\section{Conclusions}
In this paper, we introduce MM-KWS, a novel approach for user-defined keyword spotting that leverages both text and speech modalities for multi-modal prompts. By employing several multilingual pre-trained models, MM-KWS showcases its adaptability across Mandarin and English datasets. Moreover, incorporating advanced data augmentation tools for hard case mining substantially improves MM-KWS's capacity to distinguish confusable keywords. Moving forward, we plan to extend this approach into a unified single-model multilingual framework and streamline its deployment on end-side devices by prioritizing model lightweighting, aiming to validate its efficacy in real-world scenarios.

\clearpage
\section{Acknowledgements}
This study was supported by the National Natural Science Foundation of China (NSFC) under Grants 61871262, 61901251, and 62071284, the Innovation Program of Shanghai Municipal Science and Technology Commission under Grants 21ZR1422400, 20JC1416400 and 20511106603, Pudong New Area Science \& Technology Development Fund, Key-Area Research and Development Program of Guangdong Province under Grant 2020B0101130012, and Foshan Science and Technology Innovation Team Project under Grant FS0AA-KJ919-4402-0060.

\bibliographystyle{IEEEtran}
\bibliography{ref}

\end{document}